\documentclass[aps,prd,showpacs,nofootinbib,superscriptaddress,preprint,tightenlines]{revtex4}
\usepackage{graphicx}
\usepackage{amsbsy}

\usepackage{amsfonts}

\usepackage{amssymb}

\usepackage{amsmath}
\newcommand{\hateq}{\widehat{=}}
\def\be{\nopagebreak[3]\begin{equation}}
\def\ee{\end{equation}}
\def\ba{\nopagebreak[3]\begin{eqnarray}}
\def\ea{\end{eqnarray}}

\def\d{{\rm d}}

\def\f{\frac}

\def\lp{\ell_{\rm Pl}}

\def\U(1){{\rm U(1)}}

\def\d{{\rm d}}

\def\bar{\overline}

\def\={\hateq}
\def\SU(2){\rm SU(2)}
\def\U(1){\rm U(1)}

\newcommand{\teta}{\rlap{\lower2ex\hbox{$\,\tilde{}$}}\eta{}}

\newcommand{\smu}{\sin{(\bar \mu_1 c_1)}}
\newcommand{\cmu}{\cos{(\bar \mu_1 c_1)}}
\newcommand{\smd}{\sin{(\bar \mu_2 c_2)}}
\newcommand{\cmd}{\cos{(\bar \mu_2 c_2)}}
\newcommand{\smt}{\sin{(\bar \mu_3 c_3)}}
\newcommand{\cmt}{\cos{(\bar \mu_3 c_3)}}
\newcommand{\heff}{{\cal H}_{\mathrm{eff}}}
\newcommand{\rcr}{\rho_{\mathrm{crit}}}

\begin{document}
\preprint{\vbox{\baselineskip=12pt \rightline{IGC-09/5-3} }}
\preprint{\vbox{\baselineskip=12pt \rightline{PI-QG-134} }}
%\title{Singularity resolution in loop quantum cosmology: Some insights and lessons}
\title{A geometric perspective on singularity resolution and uniqueness in loop quantum cosmology}
\author{Alejandro Corichi}\email{corichi@matmor.unam.mx}
\affiliation{Instituto de Matem\'aticas, Unidad Morelia,
Universidad Nacional Aut\'onoma de M\'exico, UNAM-Campus Morelia,
A. Postal 61-3, Morelia, Michoac\'an 58090, Mexico}
\affiliation{Center for Fundamental Theory, Institute for
Gravitation and the Cosmos, Pennsylvania State University,
University Park PA 16802, USA}
\author{Parampreet Singh}
\email{psingh@perimeterinstitute.ca} \affiliation{Perimeter
Institute for Theoretical Physics, 31 Caroline Street North,
Waterloo, Ontario N2L 2Y5, Canada}

\begin{abstract}
We re-examine the issue of singularity resolution in homogeneous loop quantum 
cosmology from the 
perspective of geometrical entities such as expansion rate and the shear scalar. These quantities are very reliable measures of the properties of spacetime and can be defined not only at the classical and effective level, but also at an operator level in the quantum theory. From their behavior in the effective constraint surface and in the effective loop quantum spacetime, we show that one can severely restrict the ambiguities in regularization of the quantum constraint and rule out unphysical choices. We analyze this 
in the flat isotropic model and the Bianchi-I spacetimes. In the former case we show that the expansion rate is absolutely bounded only for the so called improved  quantization, a result which synergizes with uniqueness of this quantization as proved earlier. Surprisingly, for the Bianchi-I spacetime, we show that out of the available choices, the expansion rate and shear are bounded for only one regularization of the quantum constraint. It turns out that only for this choice, the theory exhibits quantum gravity corrections at a unique scale, and is physically viable. 
\end{abstract}

\pacs{04.60.Pp, 04.60.Ds, 04.60.Nc} \maketitle

\section{Introduction}

One of the main challenges that faces a potential quantum theory of gravity is to
resolve the classical singularities that are present in general relativity (GR), and in particular the initial singularity, or the big bang. Loop quantum gravity \cite{lqg}, 
is a candidate for such a theory. In particular, when
loop quantization methods are applied to isotropic homogeneous cosmologies, in what is known as loop quantum cosmology (LQC) \cite{lqc} the resulting theory has shown interesting promise
in this enterprise. For, when models with simple matter content
are considered, LQC yields a well defined quantum theory
for which the big bang/big crunch
gets replaced by a quantum bounce. These results first derived for a spatially flat
FRW universe with a massless scalar field \cite{aps0,aps2,polish} were generalized to include the $k=\pm1$ cases \cite{closed,open}, a cosmological constant \cite{vp} and also a $\phi^2$ inflationary potential \cite{massive}. Within the $k=0$ Robertson-Walker model and massless scalar field, the 
model could be solved exactly by choosing volume as the lapse \cite{slqc}. This enabled 
us to prove robustness of various results which were found previously through numerical simulations. These include: (i) quantum bounce for a dense subspace in the physical Hilbert space and  (ii) a supremum on the expectation values of the energy density operator which coincides with the critical energy density at which the bounce occurs. Further, using this model it is possible to gain insights on the nature of the universe before the quantum bounce. It turns out that a universe which is peaked on a classical trajectory at late times after the bounce is constrained by the properties of the quantum constraint and observables to be peaked on a classical trajectory before the bounce at early times \cite{recall}.  
For an up to date review of these results see \cite{AA:GRG}.

It is well known that a process of quantization may result in various quantization ambiguities. More importantly, regularization of the constraints can be performed in inequivalent ways of which not all may lead to a physically consistent and viable theory. One thus requires a criterion to eliminate the choices which are unphysical.
For the  $k=0$ isotropic model these criteria were put forward in Ref. \cite{CS:unique}, where 
different loop quantizations were compared and put to test by requiring a set of reasonable consistency conditions. 
It was shown that there is only one possible regularization of the constraint operator for which the following conditions are satisfied \cite{CS:unique}: (i) 
independence from the freedom underlying fiducial structures, (ii) a well defined 
critical energy density at which bounce occurs and (iii) that at small spacetime curvatures the theory should approximate GR. The only loop quantization in the flat isotropic case which satisfies these requirement is the so called `improved quantization'  \cite{aps2}.

In these studies the use of effective dynamical  equations turns out to be very helpful. Effective equations are ordinary Hamilton equations of motion found by means of an effective Hamiltonian ${\cal H}_{\rm eff}$. This Hamiltonian (that depends explicitly on $\hbar\,$ so it can not be seen as purely `classical') has been found using geometrical methods in quantum mechanics \cite{vt} and has been shown to be responsible for the dominant term to the canonical path integral of the theory \cite{path-integral}. In a recent systematic study of the resulting `effective spacetimes', it was shown that, for the $k=0$ case, they are always non-singular and there can exist no strong curvature singularities \cite{lqc-singular}. Thus, in a precise sense and for $k=0$ isotropic case, loop quantum cosmology yields a generic singularity resolution.

Even when these results are rather encouraging towards the ultimate goal of proving
that quantum effects generically resolve the classical singularity, one still needs to
overcome several challenges. For instance, when considering anisotropic models, like 
a Kantowski-Sachs and Bianchi I cosmologies, the notions that were rather useful for isotropic models already show some limitations. 
When we depart from the isotropic models, the first difficulty is that the curvature
tensor not only has a Ricci component but also its Weyl part is non-zero. This means that
the classical spacetimes can develop `Weyl singularities', in which the curvature invariants build from the Weyl part of the curvature blow up (while allowing for the Ricci part to remain well behaved). Thus even if quantum effects bind the 
Ricci scalar, the spacetime may still contain physical singularities.\footnote{The situation in the anisotropic models is thus more complex than the isotropic model where a 
bound on the energy density of the matter field translated to the bound on spacetime curvature for all matter (except whose pressure diverges at finite density) and the only singularities which may exist are harmless weak curvature type \cite{lqc-singular}. In the anisotropic model, the bound on energy density thus captures little about the fate of singularity.}
These considerations bring to the forefront the need for an unifying principle that allows us to approach the issue of generic singularity resolution in the quantum theory. In other words, even at this level one is pressed with the question of what criteria should one use to say that singularities are resolved by the quantum theory.

The question in itself poses its own problems. Singularities in the classical theory are understood in terms of concepts such as the inextendibility of 
geodesics on the spacetime
\cite{singularities}. However, the mere notion of geodesics assumes the existence of a sufficiently differentiable metric for all points on the manifold, including those that are in the region where `quantum effects' should dominate. Quantum gravity is expected to introduce some `fuzziness' near the singularity, which itself means the notion of spacetime, and geodesics, should stop to be valid and be replaced by some `quantum foam'.
How can we distinguish then
the situation where singularities are resolved from those where they are not, if the quantum state is not `semiclassical' in the strong gravity regime? In other words, is the singularity resolved just because
the state is not semiclassical near the would-be singularity? Several proposals have been put forward in this direction. These include the idea of singularity resolution via quantum foam effects \cite{wheeler,narlikar-paddy}. For a recent review of different such scenarios, from the perspective of Wheeler-DeWitt theory, see for instance \cite{kiefer}. From our viewpoint this question can only be meaningfully addressed via physical observables. This implies, in the usual language of constrained systems, analyzing the properties of Dirac observables. It is from 
this perspective that singularity resolution was established for the FRW models in LQC \cite{aps2,closed,open,slqc,warsaw}, when the relational observable defined by the matter energy density `at internal time $T$' was shown to be absolutely bounded.

The purpose of this paper is to revisit the problem of singularity resolution for homogeneous cosmologies from a geometric perspective. 
Here we put forward the viewpoint that the relevant quantities to consider are invariant geometrical quantities that are in the classical setting responsible to define when singularities arise. In particular, we shall pay close attention to well motivated scalars associated to the evolution of cosmological observers such as the expansion parameter and
the shear\footnote{As is well known, expansion and shear play a prominent role in the classical singularity theorems \cite{singularities}.}. As we shall argue, these geometrical quantities not only are useful for considering the evolution as given by the effective Hamiltonian, but
can indeed be extended to the quantum realm where the classical interpretation might not be valid. This allows for a quantum version of singularity resolution. 

Another aspect that we shall consider here concerns the issue of choosing a unique quantization for the constraint operator of the theory. In the case of isotropic
cosmologies, mathematical and physical consistency condition uniquely selects a 
quantization \cite{CS:unique}. A natural question is whether this quantization  also gets selected from the geometric perspective of singularity resolution. As we shall see, in the case of isotropic cosmologies, the answer is in the affirmative. The quantization which is naturally selected is the one resulting from the `improved quantization' of LQC \cite{aps2,slqc}. With this in mind, we analyze two quantizations of the Bianchi I model  proposed in literature, for which one expects to recover the isotropic sector. Both these quantizations are motivated by a generalization to anisotropy of the unique `$\bar{\mu}$' quantization of the
isotropic sector, but are inequivalent. As we show here, requiring that the theory has a well defined Planck scale, and therefore resolves the singularity from our geometric perspective, picks out one of these quantizations. This is the quantization which has been recently developed in detail by Ashtekar and Wilson-Ewing \cite{aa-we}. Our analysis thus leads to a lesson for the anisotropic model, which is: even if inequivalent quantizations might be expected to lead to the correct isotropic limit, a detailed analysis shows that this might not be so. Therefore, not all of them are viable.  In fact, for the Bianchi-I cosmologies our results point towards a similar sort of uniqueness as for the isotropic LQC.

The structure of the paper is as follows. In Sec.~\ref{sec:2} we give some preliminaries regarding the behavior of cosmological observers within homogeneous cosmologies, including the modifications brought in by the effective dynamics of LQC. In Sec.~\ref{sec:3} we review the isotropic FRW model from this perspective and show that the `improved' quantization in tailored to bound the expansion of the geodesics of cosmological observers and therefore gets selected as a viable choice. Section~\ref{sec:4} is devoted to the Bianchi I
model for which we compare two existing quantizations in view of our geometrical considerations. We end with a discussion and an outlook in Sec.~\ref{sec:5}.

\section{Homogeneous cosmologies and observers}
\label{sec:2}

With the purpose of having a self-contained exposition, in this section we recall some basic notions needed to treat cosmological observers in homogeneous settings. We include a discussion of the corresponding modifications that emerge from the effective dynamics of loop quantum cosmology for the isotropic case.

Let us first recall the main features of the congruence of cosmological observers and the way their expansion is related to other geometric invariants.
If $\xi^a$ denotes the congruence of  cosmological observers, then the quantity 
\be
B_{ab}:=\nabla_{b}\xi_a\, ,
\ee
is symmetric and corresponds to the 
Lie derivative of the metric in the normal direction (and for $N=1$ this is just
the proper time derivative), and also to the extrinsic curvature
$K_{ab}$ of the slicing by homogeneous spaces. The trace of this quantity is 
\be
\theta:=g^{ab}\,B_{ab}=K ~.
\ee 
That is, the expansion of the
congruence of cosmological observers corresponds to the trace 
of the extrinsic curvature.  Given that
the function $\theta$ contains the information about the expansion of the cosmological
observers, it is a natural measure of when singularities form,
namely when the expansion blows up. The evolution equation for the expansion rate $\theta$ relates it to 
the shear scalar $\sigma^2$ and the spacetime curvature:
\be \label{genray} \frac{{\rm d}\theta}{{\rm d}\tau}=-\frac{1}{3}\,
\theta^2-\sigma^{ab}\sigma_{ab}-R_{ab}\xi^a\,\xi^b
\ee
where the contribution from the term corresponding to vorticity vanishes due to the existence
of homogeneous slices. This equations holds in general for homogeneous cosmologies for which a preferred set of observers form the congruence $\xi^a$. 

For  classical GR, if matter satisfies the strong energy condition then 
$R_{ab}\xi^a\,\xi^b\geq 0$, and hence the expansion rate of an initially converging congruence diverges in 
a finite proper time. 
The classical constraint (density) relates the extrinsic curvature $K_{ab}$, the spatial Ricci scalar (which is zero in our case) and the energy density $(\rho)$,
\be\label{cc}
K^{ab}K_{ab}-K^2=16\pi G\,\rho
\ee 
which can be rewritten in terms of $\theta$ and $\sigma^2$ as
\be\label{classicalthetasigma}
16\pi G\,\rho=\frac{2}{3}\;\theta^2-\sigma^2 ~.
\ee
For the isotropic spacetime, the divergence in expansion rate or the caustic is equivalent to the divergence of energy density and also that of curvature invariants such as Ricci scalar:
\be
R = 8 \pi G  \, (1 - 3 w) \rho ~.
\ee
Here $w$ is the equation of state of the matter source: $w = P/\rho$, where $P$ denotes the pressure of the matter content. For the anisotropic spacetime, a caustic may be formed from divergence either of energy density and/or the shear scalar. In the former case the curvature singularity is Ricci curvature type and in the latter it is Weyl curvature type.

If we now consider the Hamiltonian theory, Eq.~(\ref{classicalthetasigma}) can be
rewritten  in terms of an arbitrary choice of canonical variables, denoted by $(p,q)$. 
%for simplicity and to label arbitrary canonical variables. 
Let us represent these function by $\tilde{\theta}=\tilde{\theta}(p,q)$ and $\tilde{\sigma}^2=\tilde{\sigma}^2(p,q)$. It is important to stress that these functional relations are already using some dynamical information (relating for instance, extrinsic curvature with time derivative of the three-metric, which are part of Einstein's equations in canonical form).

In the effective dynamics of LQC though Eq.(\ref{genray}) holds, the relationship between expansion $\tilde{\theta}$, shear $\tilde{\sigma}^2$ and $\rho$ is modified because the classical constraint (\ref{cc}) is no longer valid. In loop quantization, the classical constraint in terms of Ashtekar variables i.e. the connection $A$ and triad $E$ is first  expressed in terms of holonomies of connection and flux of triads and then quantized. 
For the flat 
isotropic model and the Bianchi-I spacetime, the only modifications to the classical constraint arise in the gravitational part.\footnote{For models with non-vanishing intrinsic curvature, modifications to the 
inverse scale factor operator affect matter energy density. If the intrinsic curvature vanishes, such modifications have little physical meaning \cite{aps2}.} It implies that eq.(\ref{classicalthetasigma}) the scalar constraint (and therefore, Einstein's equations) is modified as
\be
16 \pi G \, \rho = f(\tilde \theta, \tilde \sigma) ~,
\ee
%
%Here $\tilde \theta$ and $\tilde \sigma$ correspond to the classical expressions for the %expansion rate and shear scalar as defined by the classical equations.
where $f(\tilde \theta, \tilde \sigma)$ is a function determined by the quantization
prescription. 
In the effective dynamics, in order to recover general relativity, one expects that $f(\tilde{\theta},\tilde{\sigma}) \mapsto \frac{2}{3}\;\tilde{\theta}^2-\tilde{\sigma}^2$ in the appropriate regime (when $\tilde{\theta}^2,\tilde{\sigma}^2 \ll 1/l^2_{\rm Pl}$). If this is the case then we expect that the dynamical equations
associated to the effective Hamiltonian will approximate Einstein's equations also in this regime.

In order to illustrate the way key equations get modified, let us consider the case of a
$k=0$ Robertson-Walker metric where all expressions can be explicitly computed (recall that in this case $\tilde{\sigma}^2=0$). In this case, the loop quantum effective Hamiltonian for the `improved quantization', when described in canonical variables $(\beta,V)$, is given by  \cite{aps2, slqc, CS:unique}
\be\label{const3}
{\cal H}_{\rm eff} = - \frac{3}{8\pi G\gamma^2}\,V\, \frac{\sin^2(\lambda\,\beta)}{\lambda^2} + \rho\,V ~.
\ee
where $V$ denotes the physical volume and $\lambda = 4 \sqrt{3} \pi \gamma \lp^2$ with $\gamma = 0.2375$ \cite{aa-we}. The value of $\lambda$ is determined by the procedure of regularization of the Hamiltonian constraint in LQC.

Thus, for LQC, the function $f(\tilde \theta, \tilde \sigma)$ in the flat isotropic model is
\be
f(\tilde{\theta}):= 6 \, \frac{\sin^2(\lambda\,\beta)}{\gamma^2\,\lambda^2}
\ee
where $\beta$ is  canonically conjugate to the volume $V$ and satisfies $\{ \beta, V\}= 4\pi G\gamma$. (On the classical solutions $\beta$ is related to the scale factor as $\beta = \gamma \dot a /a$).
The quantum gravity parameter  $\lambda$ has dimensions of length and is of the order of Planck length. 
This parameter fixes the scale near which quantum gravity become significant.
In fact, when $\lambda\,\beta \ll 1$,
we recover the classical constraint $\tilde{\theta}=\beta^2/\gamma^2=\frac{8\pi G}{3}\rho$. 

In order to evaluate the expansion factor in the effective LQC we first use Hamilton's equation:
$\dot{V}=\{ V, {\cal H}\}$ and then compute 
$\theta:=\dot{V}/V$. 
For the effective Hamiltonian ${\cal H}_{\rm eff}$, we get,
\be
\theta:=\frac{1}{V}\;\{ V, {\cal H}_{\rm eff}\}=\frac{3}{\lambda\gamma}\sin(\lambda\beta)\cos(\lambda\beta)=\frac{3}{2\lambda\gamma}\sin(2\,\lambda\beta)
\ee
Note that we have retained the use of $\theta$ to denote the expansion as seen by the cosmological observers, in the underlying spacetime.
In the classical limit, $\lambda\mapsto 0$, we recover the relation 
$\theta \approx 3\beta/\gamma = \tilde{\theta}$.\footnote{Note also that even when 	$\theta \approx 3\beta/\gamma$ when $\lambda\beta \ll  1$, it is different from the expression $\tilde{\theta}':=3\sin(\lambda\beta)/\gamma\lambda$ as it appears in the constraint (\ref{const3}). This difference between the quantity that replaces $\tilde{\theta}$ 
in the modified constraint (\ref{const3}), and the expansion as computed using the dynamical equations, is what allows to have an `effective Friedman equation'. This is qualitatively different to the classical equation (equivalent to the classical constraint (\ref{classicalthetasigma})), when written in terms of density, by using the identity $\sin^2(2\lambda\beta)/4=\sin^2(\lambda\beta)(1-\sin^2(\lambda\beta))$.} 
Using the
effective constraint (\ref{const3}) along with above Hamilton's equation we get
\be
H^2= \frac{\theta^2}{9} = \frac{8\pi G}{3} \rho \;\left(1-\frac{\rho}{\rcr}\right)
\ee
with $\rcr = 0.41 \rho_{\mathrm{Planck}}$ the critical density for which the Hubble parameter vanishes $H=0$ and a bounce occurs. At classical scales i.e.  $\rho \ll \rcr$, $\theta \rightarrow \tilde{\theta}$ and the above equation approximates eq.(\ref{classicalthetasigma}) for the isotropic model.

Using the Hamilton's equation for $\beta$, it is straightforward to evaluate 
$\dot \theta + \theta^2/3 = 3 \ddot a/a$ and find, from Eq.~(\ref{genray})
\be
R_{ab} \, \xi^a \xi^b = 4 \pi G \rho  \left(1 - 4\,\f{\rho}{\rcr}\right) + 12 \pi G P\, \left(1 - 2\, \f{\rho}{\rcr}\right) ~.
\ee
This implies that in effective dynamics of LQC, $R_{ab} \xi^a \xi^b$ can become negative near the Planck scale, even for matter which satisfies the strong energy condition. 
It is also illustrative to rewrite
Raychauhuri equation as,
\be
\dot{\theta}=-12\pi\,G\,(\rho + P)\,\left(1 -2\,\f{\rho}{\rcr}\right)\, .
\ee 
As an example, for the massless scalar field, where $(\rho = P)$, the quantity  $R_{ab} \, \xi^a \xi^b$ becomes negative when $\rho > 2/5\, \rcr$. Since $R_{ab} \xi^a \xi^b \geq 0$ is violated in this regime, there is a repulsive force that out-weights the gravitational attraction, in such a way that the expansion reaches
its maximum absolute value when $\rho = 1/2\, \rcr$ and then $|\,\theta|$ begins to decrease
until it reaches zero at the bounce, when $\rho = \rcr$. Note also that this qualitative behavior of the expansion is valid when the matter satisfies the null energy condition.
Therefore, caustics do not form in the effective spacetime of flat and isotropic LQC. In view of this, it is not surprising that this spacetime is devoid of {\it  any} physical singularities \cite{lqc-singular}.

\section{Isotropic $k=0$ FRW}
\label{sec:3}

The flat, isotropic and homogeneous Robertson-Walker metric is one of the simplest cosmological model which can be quantized in LQC. Even though 
simple, this model offers useful insights on the regularization of the Hamiltonian constraint and restrictions on quantum ambiguities for models with more degrees of freedom and even the full theory. It is well known that out of the available 
``loop quantizations'' for this model, only the so called ``improved quantization'' \cite{aps2} leads to a physically viable description. Unlike other regularizations of the 
Hamiltonian constraint, the one described by the $(\beta, V)$ variables (as in Sec.~\ref{sec:2}) leads to 
a well defined Planck scale and approximates GR at low spacetime curvature for matter satisfying the null energy condition \cite{CS:unique}.

In the following we re-analyze the quantization of the flat FRW model for
different `loop quantizations' and consider the behavior of the corresponding
operator for the expansion rate of cosmological observers 
in all these theories. Perhaps not surprisingly we find that, even when many
such quantizations might display a `bounce', there is a unique choice for which
the corresponding operator is bounded, and therefore, when interpreted in
terms of the effective description, the resulting bounce occurs at a unique
`Planck scale'. This choice corresponds to the improved quantization of \cite{aps2}. In all other choices, including the so called `$\mu_0$
quantization' (and all lattice refining models \cite{lattice}), the corresponding
operator is not bounded. As expected, the quantum gravity scale in this model turns out to be arbitrary and the theory lacks any predictive power.  In this case,
when the effective description is studied,
the maximum value of the expansion (or acceleration) a geodesic congruence encounters can
be arbitrarily low, well below the `Planck scale' $\theta \approx 1/l_{\rm
Pl}$. One is then led to conclude that there is a unique quantization prescription
for which singularity resolution is well defined {\it on the full space of states} 
of the theory. We find
therefore an interesting interplay with the results of \cite{CS:unique}.

Let us recall the most basic facts about the $k=0$ models. In the ordinary
geometrodynamical
description, the geometric configuration variable is the scale factor $a$ and
its corresponding conjugate momenta $P_a= - 3 V_o \, a \,\dot a/(4 \pi G)$,
with $V_0$ the fiducial volume of the region under consideration (for
simplicity we shall consider a closed
$\mathbb{T}^3$ topology for the spatial manifold).\footnote{The results which follow are not affected by any other choice including that of non-compact $\mathbb{R}^3$ manifold. In the latter case the following discussion is supplemented with subtleties of the freedom of the rescaling of the fiducial cell which is discussed in detail else where \cite{CS:unique}.}
When using the basics variables that are employed
in loop quantum gravity, namely, densitized triads and connections, it is
convenient to consider the variables $|p| = V_o^{2/3} \, a^2$ and $c = \gamma
\, V_o^{1/3}\, {\dot a}$, valid on solutions, for which we have canonically
conjugate variables $(c,p)$, satisfying
\be
\{c,p\} = \f{8 \pi G \gamma}{3} ~.
\ee
Starting from these variables, one can consider a rather general possibility
of variables,
\be\label{genvar}
P_g = c \, p^m \,, \qquad g = \f{p^{(1 - m)}}{1 - m}
\ee
obtained from $(c,p)$ by a canonical transformation, for which the
effective Hamiltonian constraint becomes \cite{CS:unique}
\be
\f{3}{8 \pi G} \f{\sin^2(\lambda_{{P_g}} P_g)}{\gamma^2  \lambda^2_{P_{g}}} \,
\left((1 - m) g\right)^{(1 - 4 m)/(2(1 - m))} = H_{\mathrm{matt}} ~
\ee
From here one can compute the expansion rate $\theta=\dot{V}/V$,
and get
\ba
\theta &=& \frac{3}{2(1-m)}\frac{\dot{g}}{g}  \nonumber \\
& =& \frac{9}{8\pi}
\frac{(1-m)^{-4m}}{\gamma^2\lambda_{P_g}}\sin(\lambda_{{P_g}} P_g)
\cos(\lambda_{{P_g}} P_g)\, g^{-\frac{(2m-1)}{2(1-m)}} ~.\nonumber
\ea
It is now straightforward to see that the absolute value of the expansion will
reach a
maximum value $|\theta|_{\mathrm{max}}$ before decreasing and becoming zero (when
$\cos(\lambda_{{P_g}} P_g)$ vanishes and the bounce occurs).
It is also easy to see that the maximum value, on a given trajectory, also 
depends on the phase space variables. In particular, the value of $|\theta|_{\rm
max}$ is determined by the initial conditions for the matter content except 
for one case. This is for the choice $m=-1/2$, which  yields $P_g=\beta$ and $g=2V/3$, and
therefore corresponds to
the `improved $\bar{\mu}$ quantization' \cite{aps2,CS:unique}. Only in this case 
$|\theta|$ is absolutely bounded, on the constrained surface, by a phase space independent number $|\,\theta|_{\rm sup}$.

In order to appreciate the significance of this result, let us consider a
particular example, namely the so called `$\mu_0$' quantization with a
massless scalar field (for a discussion of other inadequacies of this
quantization see \cite{aps2,CS:unique}). In this case $m=0$, and
$|\theta|_{\rm max}$ is finite for any trajectory, but it depends on $P_\phi$,
a Dirac observable that labels the trajectory on phase space. Since $\theta$ is bounded on each trajectory (though by dependence on $P_\phi$), one may naively conclude that 
the theory achieves  singularity resolution. As we shall now argue this 
conclusion is incorrect.
This can be seen in two steps. First, if we only consider the effective
equations, and pretended that there is an underlying spacetime at all points
of the `evolution' on phase space, the scale (with dimensions $L^{-1}$) at
which quantum gravity effects become important 
can have {\it any value}, since the scale depends on $P_\phi$ which is completely
free. This means that quantum effects can kick in at an arbitrarily small scales in contradiction to various experiments, or at an arbitrary scale well beyond the Planck scale.
Clearly this scenario is unphysical.

But, is this behavior an artifact of considering the effective equations to be
valid at all regimes? Can we forgo this pathological possibilities when we consider the
full quantum theory without going to the effective description? In other
words, can we still have
quantum singularity resolution in the `$\mu_0$ scheme'? Let us now see that,
when the quantum
theory is considered and the question is carefully analyzed, the answer is
still in the negative.
For that, let us consider in the quantum theory, for arbitrary $m$, an operator ${|\,\hat\theta|}$
associated to the expansion, where we `promote' the (effective) classical
expression to an operator in the underlying geometrical sector of the Hilbert space of the theory.
Clearly, this quantity has the standard geometrical interpretation for those
states that are semiclassical and whose dynamics approximate that of a
geometry satisfying classical equations. An important feature is that, being a
well defined operator, it can also be defined on those state with no
semiclassical interpretation. If we now compare the corresponding operators
for all possible choices of $m$, we are led to conclude that it is
only for the improved quantization that the operator is bounded in the
physical Hilbert space of the theory. Let us be more precise. The function $\theta$
that represents expansion for this particular quantization, is bounded on the constraint
surface. Therefore, the corresponding operator on the quantum theory will have a bounded
spectrum. This is precisely realized in the exact quantization of \cite{slqc}.
This result is, in this simple isotropic case,
rather similar to the behavior of energy density that was found to be
bounded in this same quantization \cite{slqc,warsaw}.

These considerations suggest the following conclusions. First, one
could have as a necessary condition for singularity resolution in the full
quantum theory that the operator ${|\,\hat\theta|}$ be a bounded operator. In
the $k=0$ case, this immediately implies that this condition will only be
satisfied by the improved quantization, pointing again to a uniqueness result
along the lines of  \cite{CS:unique}. Secondly, in the resulting effective theory,
the underlying `classical geometries' will have a well defined Planck scale as defined by the expansion of the cosmological observers, only when the condition of boundedness of $\theta$ is satisfied.

%\section{Bianchi I}
\section{Bianchi-I Model}
\label{sec:4}

The loop quantization of Bianchi-I model is an interesting generalization of the 
isotropic loop quantum cosmology. Unlike the isotropic model, the Weyl curvature is not zero in this model and hence singularities are not only of Ricci curvature type but also of Weyl curvature type. As emphasized in our discussion earlier, this 
feature makes it necessary to go beyond the scope of energy density and its properties to understand singularity resolution.

The spatial manifold in the Bianchi-I model can in general be non-compact with $\mathbb{R}^3$ topology. %If the spatial manifold is non-compact then i
Just as in the isotropic case, in order to introduce a symplectic structure, it is  necessary to consider a fiducial cell ${\cal V}$. With respect to the fiducial metric introduced on $\mathbb{R}^3$, this cell has volume $V_o = l_1 l_2 l_3$.
The size of the fiducial cell and the fiducial lengths $l_I$ are completely arbitrary freedoms which must be respected by the underlying theory at the classical as well at the effective and quantum level. As we will see the consistent implementation of these freedoms severely constrain the available quantization choices. If the spatial manifold is restricted to a compact topology 
such as a three-torus ($\mathbb{T}^3$), then the fiducial cell is no longer required as the integrations to define the symplectic structure are finite. However, we will show that even with this topology the choices which are restricted as above remain inviable.

The spacetime metric in the Bianchi-I model, for a particular choice of fiducial coordinates $(x,y,z)$ is of the form:%In the Bianchi-I model the spacetime metric is of the form
\be 
\d s^2=-N^2\,\d t^2+a_1^2\,\d x^2 + a_2^2\,\d y^2 + a_3^2\,\d z^2 
\ee 
where $N$ is the lapse factor. The metric reduces to the isotropic one when the individual scale factors become equal.\footnote{There is still the possibility of changing the fiducial coordinates that will introduce a further freedom, in which case an isotropic geometry will not have the same factors $a_I$. This fact introduces a subtlety in the passage from the Bianchi model to the isotropic one. For a detailed discussion of these issues see \cite{aa-we}.}
Due to the underlying symmetries of the spacetime, the matrix valued connection $A^i_a$ and triad $E^a_i$ can be expressed as $c_I$ and $p_I$ respectively (where $I = 1,2,3$). These phase space variables satisfy 
%canonical conjugate phase space variables satisfy
\be
\{c_I, p_J\} = 8 \pi G \gamma \delta_{IJ} ~
\ee
where $\gamma$ is the Barbero-Immirzi parameter. The metric components are related to the triads as
\be
|p_1| = l_2 l_3 \, a_2 a_3, ~~~ |p_2| = l_1 l_3 \, a_1 a_3, ~~~ |p_3| = l_2 l_3 \,a_2 a_3 ~.
\ee
On the solutions of classical GR, the connection components are related to 
the time derivative of metric components as
\be
c_I = \gamma \, l_I \, \dot a_I ~.
\ee
It is useful to note that under the freedom of the change in the shape of the fiducial cell $(l_1,l_2,l_3) \rightarrow (l_1',l_2',l_3')$, $(c_I,p_I)$
transform as 
\be 
c^1 \rightarrow c'_1 = l_1' c_1, \quad p_1
\rightarrow p_1' = l_2' l_3' p_1 ~
\ee 
and similarly for other components.

Any quantization of this model should satisfy two consistency requirements: (i) If performed for a non-compact spatial topology, it must respect above freedom as well as the one in the choice of underlying fiducial metric. This corresponds to independence of physical predictions of invariant quantities such as expansion factor and curvature scalars from $V_o$ and $l_I$ and (ii) the quantum theory and its physical implications must reduce to those of the isotropic model in loop quantum cosmology when the anisotropic shear vanishes.
The latter requirement ensures that the theory should have GR as its infra red limit at low spacetime curvatures.

In the literature two different loop quantizations have been proposed which reduce to the 
consistent isotropic case in loop quantum cosmology. Both of these choices draw lessons from 
the `improved quantization' of \cite{aps2}, albeit in a different way. The first choice relies on 
extending the prescription to regularize the field strength tensor in the anisotropic case ``as is'' in the 
isotropic case for triad variables \cite{chiou}. For the massless scalar field the resulting quantization can be shown to have certain similar features as the exactly solvable loop quantum cosmology \cite{slqc} and has been extensively studied in different contexts \cite{cv:bianchi,bianchi,lukas,madrid}. However, it turns out that 
this quantization suffers from lack of freedom for the choice of shape of the fiducial cell and one is therefore restricted to a compact topology, such as a torus. The second choice for the regularization of the Hamiltonian constraint \cite{aa-we} (see also Appendix C of Ref. \cite{cv:bianchi} where it was first discussed) successfully overcomes these limitations. At first sight it appears to be based on a prescription which is not as parallel to the improved quantization of isotropic LQC. However, this viewpoint is incorrect. As we will show, this quantization in fact completely captures the `improved dynamics' as expressed in $\beta, V$ variables. 
%Before we discuss this quantization in some detail, two remarks are in order:\\
The first and second choice of regularization of the Hamiltonian 
constraint are sometimes referred in the literature as $\bar \mu$ and $\bar \mu'$ approach to reflect 
that the first choice mimics the $\bar \mu$ or `improved quantization' of isotropic LQC. Such a 
nomenclature is misleading for two reasons. The first is that in the isotropic limit both regularizations yield the $\bar \mu$ `improved dynamics'. Secondly, as we will show below, it is in fact the $\bar \mu'$ approach which truly captures the regularization of Hamiltonian constraint as performed in isotropic LQC. However, in order for comparison of our analysis with earlier works we shall use above notation. In principle it is possible to perform a general analysis incorporating other regularizations by introducing variables analogous to those in (\ref{genvar}) for the Bianchi-I spacetime. These will include those which do not lead to the unique consistent isotropic quantum dynamics in an appropriate limit. However, such models (which include the so called lattice refined models \cite{lattice}) suffer similar limitations as for the first choice of regularization (the $\bar \mu$ scheme). These will not be considered in the following.%\\

We first discuss the quantization as proposed in Ref. \cite{aa-we}. 
As we remarked earlier, the resulting theory respects the underlying freedom of the change in shape of the fiducial cell as well the fiducial metric.
For the massless scalar field model, the quantum constraint has been shown to be non-singular for all states in the physical Hilbert space. As in the isotropic case,
this model leads to a non-singular bounce of the scale factors when the spacetime curvature becomes close to the Planck value. For states which are semi-classical at late times, the underlying quantum dynamics can be expected to be very well approximated by the following effective Hamiltonian %\footnote{This effective Hamiltonian was first proposed in ref. \cite{cv:bianchi} as the $\bar \mu'$ scheme.}
\be\label{effham}
\heff =  - ~\f{1}{8 \pi G \gamma^2 V}\left(\f{\sin(\bar \mu_1' c_1)}{\bar \mu_1'} \f{\sin(\bar \mu_2' c_2)}{\bar \mu_2'}  p_1 p_2 + \mathrm{cyclic} ~~ \mathrm{terms}\right) + {\cal H}_{\mathrm{matt}}~  
\ee
where we have chosen lapse $N = 1$ and 
\be \label{mub1}
\bar \mu_1' = \lambda \sqrt{\f{ p_1 }{p_2 p_3}}, ~~~ \bar \mu_2' = \lambda \sqrt{\f{p_2}{p_1 p_3}}, ~~~ \mathrm{and} ~\bar \mu_3' = \lambda \sqrt{\f{p_3}{p_1 p_2}} ~ .
\ee
%Here $\lambda$ is the square root of the area gap $\Delta = 4 \sqrt{3} \pi \gamma \lp^2$. 
Above particular dependence on the triads arises from the regularization procedure employed in the quantization \cite{aa-we}.

It is interesting to note that using eqs.~(\ref{mub1}) one can rewrite the Hamiltonian constraint (\ref{effham}) in the following form
\be
\heff =  \nonumber - ~\f{1}{8 \pi G \gamma^2 \lambda^2} \, V \left(\sin(\lambda \beta_1) \sin(\lambda \beta_2) + \mathrm{cyclic} ~~ \mathrm{terms}\right) + {\cal H}_{\mathrm{matt}}~ 
\ee
where $\beta_I$ are conjugate variables to the volume $V$. In the limit $\lambda \rightarrow 0$, $\beta_I \approx \gamma \dot a_I/a_I$. Note that the form of the constraint and the behavior of $\beta_I$ is  analogous to the one in the isotropic case (\ref{const3}). Thus, as in the improved quantization of the isotropic LQC, this quantization implements a regularization of $\beta_I$ which is related to the expansion rate.

An immediate consequence of (\ref{effham}) is that it leads to a universal bound on the matter energy density. The vanishing of the Hamiltonian constraint, $\heff \approx 0$ implies
\be\label{rho1}
\rho = \f{1}{8 \pi G \gamma^2 \Delta} \left(\sin(\bar \mu_1' c_1) \, \sin(\bar \mu_2' c_2)  + \mathrm{cyclic} ~~ \mathrm{terms}\right) ~.
\ee
Since the terms in the parenthesis are all bounded functions, the maximum value of the energy density is given by $\rho_{\mathrm{max}} = 3/(8 \pi G \lambda^2)$.
Using Hamilton's equation for $p_I$:
\be
\dot p_I = - 8 \pi G \gamma \f{\partial}{\partial  c_I}  \heff
\ee
we obtain
\be
\f{\dot{p}_1}{p_1}= \f{1}{\gamma\lambda}\cos(\bar\mu_1' c_1)\left(
\sin(\bar\mu_2'\,c_2)+\sin(\bar\mu_3'\,c_3)\right)
\ee
and similarly for the other terms. One can combine them and compute the
expansion $\theta$ of the congruences of cosmological observers
\ba\label{theta1}
\theta &=& \f{\dot V}{V} =   \f{1}{2} \, \left(\f{\dot p_1}{p_1} + \f{\dot p_2}{p_2} + \f{\dot p_3}{p_3} \right) \nonumber \\
&=& \f{1}{2 \gamma \lambda}\left[\sin(\bar\mu_1'\,c_1+\bar\mu_2'\,c_2)+
\sin(\bar\mu_1'\,c_1+\bar\mu_2'\,c_2) + \sin(\bar\mu_2'\,c_2+\bar\mu_3'\,c_3)\right] .
\ea 
It is then obvious that the absolute value of the expansion $|\,\theta|$
is bounded, on the constrained surface as given by the effective dynamics.
Just as in the isotropic model, it is expected that the spectrum of the corresponding operator in the quantum theory will be bounded. The maximum value of the expansion 
factor is given by 
\be |\theta|_{\rm sup}= \f{3}{2 \gamma \lambda}
\ee 
This
means that there is an invariant `Planck scale' that binds this
covariantly defined quantity and thus provides a well defined
notion of a scale that appears as a true quantum gravity effect
(note that it is inverse to the Planck length and thus would
diverge in the `classical limit' $\ell_{\rm Pl}\mapsto 0$).

Another consequence of the effective Hamiltonian $(\ref{effham})$ is that the shear scalar 
\be\label{shear}
\sigma^2 := \sigma^{ab}\sigma_{ab}=\f{1}{3} \left((H_1 - H_2)^2 + (H_2 - H_3)^2 + (H_3 - H_1)^2\right)
\ee
also has a universal bound. This can be proved by finding directional Hubble rates $H_I$ defined as the ratio $\dot a_I/a_I$. A straightforward calculation leads to 
\ba\label{hubblediff1}
H_1 - H_2 &=&  \f{1}{\gamma \lambda} \Bigg[\sin(\bar \mu_1' c_1 - \bar \mu_2' c_2) + \sin(\bar \mu_3' c_3) (\cos(\bar \mu_2' c_2) - \cos(\bar \mu_1' c_1))\Bigg],\\
H_2 - H_3 &=&  \f{1}{\gamma \lambda} \Bigg[\sin(\bar \mu_2' c_2 - \bar \mu_3' c_3) + \sin(\bar \mu_1' c_1) (\cos(\bar \mu_3' c_3) - \cos(\bar \mu_2' c_2))\Bigg],\\
H_3 - H_1 &=& \f{1}{\gamma \lambda} \Bigg[\sin(\bar \mu_3' c_3 - \bar \mu_1' c_1) +  \sin(\bar \mu_2' c_2) (\cos(\bar \mu_1' c_1) - \cos(\bar \mu_3' c_3)) \Bigg]. \ea
Since each of $(H_I - H_J)$ is bounded, it is obvious that $\sigma^2$ is also bounded.

It is important to note that for the effective Hamiltonian (\ref{effham}), both $\theta$ and $\sigma^2$ are invariant quantities with respect to any freedom under the change of fiducial cell. In terms of these quantities it is quite transparent to reduce to the isotropic case, which corresponds to dynamical trajectories for which $H_1=H_2=H_3$, and therefore, $\sigma^2=0$.
Further, the upper bounds on both the entities are universal with values close to the Planck scale. \\

Let us now consider the effective Hamiltonian which arises in the alternative quantization (the so called `$\bar \mu$' approach)\cite{chiou,cv:bianchi,bianchi,lukas,madrid}. With lapse $N = V$ it is given as
\ba
{\cal{H}}_{\mathrm{eff}}  &=& \nonumber - \f{1}{8 \pi G \gamma^2} \left(\f{\smu \smd}{\bar \mu_1 \bar \mu_2} \, p_1 p_2 + {\mathrm{cyclic}} \, {\mathrm{terms}} \right) \\ ~&& + ~ V^2 \rho
\ea
where
\be
\bar \mu_1 = \f{\lambda}{\sqrt{p_1}}, ~ \bar \mu_2 = \f{\lambda}{\sqrt{p_2}} ~~ \mathrm{and} ~~ \bar \mu_3 = \f{\lambda}{\sqrt{p_3}} ~.
\ee
%(In the literature this effective Hamiltonian is also known as the `$\bar \mu$ approach' for the Bianchi-I model).

The vanishing of the above Hamiltonian constraint leads to the expression for the energy density:
\be
\rho = \f{1}{8 \pi G \gamma^2 \lambda^2} \left(\f{\sqrt{p_1 p_2}}{p_3} \, \smu \smd +  {\mathrm{cyclic}} \, {\mathrm{terms}} \right) ~.
\ee
Even though the trigonometric terms in the parenthesis are bounded, $\rho$ fails to have an upper bound because prefactors such as $\sqrt{p_1 p_2}/p_3$ can be unbounded. As an example, if the spatial manifold is a torus with a unit volume then 
$\sqrt{p_1 p_2}/p_3 = \sqrt{a_1/a_2}$ which diverges when $a_1 \rightarrow \infty$ or $a_2 \rightarrow 0$. Both of these limits are allowed possibilities in this model. Even when these prefactors are bounded, this
bound depends on the phase space variables, as was the case for
the isotropic case for a parametrization different from the improved one. 

To find the expression for the expansion factor $\theta$, let us consider the equation of motion for $p_I$. Using Hamilton's equation
\be
\f{\d p_1}{\d \tilde t} = - 8 \pi G \gamma \, \f{\partial}{\partial c_1} \heff
\ee
and $\d \tilde t = (1/V) \d t$, we obtain
\be
\f{\d p_1}{\d t} = \f{p_1 \, \cmu}{V \gamma} \left(p_2 \smd + p_3 \smt\right) 
\ee
and similarly for $\dot p_2$ and $\dot p_3$. The expression for $\theta$ thus 
becomes
\ba
\theta 
&=& \nonumber \f{1}{2 \gamma V} \Bigg[p_1 \, \smu (\cmd + \cmt)  + p_2 \, \smd (\cmu + \cmt) \\ && \hskip1cm +~~  p_3 \, \smt (\cmu + \cmd) \Bigg] \, .
\ea
It is obvious that unlike the expansion factor in Eq.(\ref{theta1}), this expression has no upper bound. Especially, the expansion factor diverges when either of the scale factors vanish (a behavior which is in common with the classical theory). On the other hand using (\ref{theta1}) we note that for loop quantization of Bianchi-I spacetime as developed by Ashtekar and Wilson-Ewing \cite{aa-we} the expansion factor {\it never} diverges even when the
volume vanishes.

The shear scalar (\ref{shear}) shares the same fate as the expansion factor in this scheme. To see that we evaluate the directional Hubble rates from $\dot p_I$ and find
\ba
H_1 - H_2 &=& \nonumber \f{1}{\gamma V} \Bigg[p_3 \, \smt (\cmd - \cmu) +   p_1 \, \smu \cmd  \\ && \hskip1cm - ~~ p_2 \, \smd \cmu  \Bigg], \\
H_2 - H_3 &=& \nonumber  \f{1}{\gamma V} \Bigg[p_1 \, \smu (\cmt - \cmd)  +   p_2 \, \smd \cmt  \\ && \hskip1cm - ~~ p_3 \, \smt \cmd  \Bigg], \\
H_3 - H_1 &=&   \nonumber \f{1}{\gamma V} \Bigg[p_2 \, \smd (\cmu - \cmt)  +   p_3 \, \smt \cmu   \\ && \hskip1cm - ~~ p_1 \, \smu \cmt  \Bigg] ~. 
\ea
Thus unlike the difference of Hubble rates in (\ref{hubblediff1}), above expressions are unbounded implying that the shear scalar in loop quantization of Bianchi-I model \cite{chiou, bianchi} may in general diverge for arbitrary matter content.

This quantization prescription then shares the same properties that we analyzed in Sec.~\ref{sec:3} for the `$\mu_0$ quantization' (and lattice refining models). That is, even when there might be
solutions to the effective equations of motion that `bounce', there is no universal `quantum gravity scale'.
In the quantum theory this inadequacy manifests itself in the fact that the operators associated to $\theta$ and $\sigma^2$ will not be bounded. Just as we did for the isotropic case, we are led to conclude that this quantization, be it on ${\mathbb R}^3$ or on the three-torus ${\mathbb T}^3$, is physically inviable.

In summary, we have analyzed the effective theory one
expects to get from two possible quantizations of the Bianchi-I
system. We saw that the so-called $\bar\mu$ quantization does not
yield a unique scale for the expansion and shear, whereas the $\bar\mu'$
quantization does.  
%We also saw that the $\bar\mu'$ quantization
%also has a bounded shear, whereas the other choice does not.
Just as we did for the isotropic case, one can
expect that the existence of bounded operators and therefore
an invariant Planck scale \cite{CS:unique}, formulated in terms of 
the expansion and shear, can be taken as criteria for selecting the physically correct
quantization.

\section{Discussion and Outlook}
\label{sec:5}

Let us summarize our results.  We have 
put forward the proposal that, in order to address the issue of singularity resolution in {\it any} quantum theory of gravity, the natural geometrical quantities to consider for
exploring the resulting quantum geometry are those quantities
associated to some observers. In the case of homogeneous cosmologies
such natural quantities are given by the expansion and shear as measured by the cosmological comoving observers. The advantage of considering these observables is twofold. First, they posses a clear geometrical interpretation when there is an underlying classical spacetime, in terms of the behavior of the congruence of geodesics associated to cosmological observers. 
Second, since they are well defined quantities, independent 
of any fiducial structures, they can also represent meaningful relational observables for the full quantum theory. They are the most natural with respect to which singularity resolution can be posed. Geodesics will be well defined and complete if these quantities
are well behaved classically. Thus, one can expect that, if the quantum geometry is non-singular, the corresponding operators are to be bounded in the quantum theory. This also represent a step further from considering only the behavior of geodesics in an effective spacetime description as was done in \cite{lqc-singular}.

With these geometrical considerations in mind, we have revisited the loop quantization of $k=0$
FRW and Bianchi I cosmologies and focused on the issue of singularity resolution.
A detailed study of these models lead us to argue that boundedness of these quantities in the quantum theory is a consistent requirement for singularity resolution, in our present restricted homogeneous context. We have seen that, if this condition is violated, as it happens in certain `loop quantizations' for certain choices in the definition of the quantum theory, the resulting effective theory has the unphysical feature of not having a well defined `quantum gravity scale'. That is, the classical trajectories on phase space might represent universes that `bounce', but such that the scale at which this happens is arbitrary, even at scales of ordinary phenomena, already ruled out by experiments. 
Another lesson we can learn from this
result is that we can, in retrospect, see {\it why} the `improved quantization' for the flat isotropic model is the only physically viable choice. It is `designed' to always bind the expansion, and therefore allow for geodesics --in the semiclassical scenario-- to be complete. As we showed, this conclusion also holds true for the expansion rate and the shear scalar in Bianchi-I model for the $\bar{\mu}'$ quantization studied in Ref. \cite{aa-we}. Even though a quantization inequivalent to the one in Ref.\cite{aa-we} can be  performed such that the theory leads to a correct isotropic limit \cite{chiou,bianchi,lukas,madrid}, our analysis shows that it fails to be viable from the geometric perspective of singularity resolution. %In the latter and other alternate quantizations, expansion rate and shear scalar turn out to be unbounded making the theory inviable. 
It is thus interesting that the uniqueness result for the isotropic model established in Ref. \cite{CS:unique} and in this paper (from a different perspective) seems to extend to the Bianchi-I spacetime.

The criteria that we have put forward allow us to speak meaningfully about singularity resolution in the isotropic case, and extend it to the anisotropic scenario.     
In both these cases, out of the available quantizations, there is only one choice which leads to a  physically viable resolution of singularities.  Thus, the requirement of having a singularity free theory selects, in these cases, a unique consistent quantization. Still, even when one can be reasonably confident that this criteria is better motivated than the previous one available for the isotropic models \cite{CS:unique}, more work is needed to assure that it will continue to be the most reasonable condition for more general models such as inhomogeneous cosmologies. 

In this respect it is perhaps useful to end this contribution with a discussion regarding the issue of singularity resolution in quantum gravity. While it is generally expected that a quantum theory of gravity will `cure' the singularities that are present in the general theory of relativity, it is not clear exactly {\it how} this is supposed to take place. 
Even in the restricted setting of homogeneous cosmologies, where only gravitational configurations with a large degree of symmetry are considered, this issue is not fully settled. For instance, if we take the viewpoint that in the classical theory (almost) all solutions possess a singularity, and if these classical solutions are to represent the phase space of the theory (as say, in geometric quantization) or be the dominant ones in the calculation of the path integral, then the resolution of the singularity by quantum effects seems somewhat puzzling. As is not so uncommon, reasonable expectations do not get realized when attempting to build detailed models for fundamental physics. Loop quantum cosmology has given an answer that is somewhat unexpected. The way that the singularity is resolved is by changing drastically that set of `classical configurations' that define the `quantum phase space'. For, all solutions to the effective equations of motion are nonsingular, and those are precisely the trajectories that dominate the canonical path integral. A second point worth emphasizing is the way that singularities are resolved in LQC. It is not through the behavior of the wavefunction on the `would-be singularity', nor through special boundary/initial conditions, but rather by analyzing the behavior of relevant Dirac observables. From this perspective, the main challenge is to identify those observables
that are relevant in order to claim that singularities are generically resolved. As we have seen, in the homogeneous setting the choice is rather natural. Whether the criteria proposed here will still be meaningful when considering inhomogeneous cosmologies remains an open issue.

\section*{Acknowledgments}

\noindent
We would like to thank Abhay Ashtekar and Guillermo Mena for discussions.
This work was in part supported by CONACyT U47857-F
grant, by NSF PHY04-56913 and by the Eberly Research Funds of Penn
State.  Research at Perimeter Institute is supported
by the Government of Canada through Industry Canada and by the
Province of Ontario through the Ministry of Research \&
Innovation.

\end{document}